# THE KINETIC OF THE ATOMIC RELAXATION INDUCED BY LASER NOISE

O. El Akramine[♣] and A. Makhoute[♣,*]

[♣] UFR Physique Atomique, Moléculaire & Optique Appliquée
Université Moulay Ismail, Faculté des Sciences,
B.P. 4010 Beni M'hamed, Meknès, Morocco.

[*] Physique Atomique Théorique, Faculté des Sciences,
Université Libre de Bruxelles,
CP 227, Brussels, Belgium.

# Abstract

We present a theoretical study of strong laser-atom interactions, when the laser field parameters are subjected to random processes. The atom is modelled by a two–level and three–level systems, while the statistical fluctuations of the laser field are described by a pre-Gaussian model. The interaction of the laser–target is treated nonperturbatively by using the calculation method based on the hermitian Floquet theory. Our aim consists in studying the kinetic of atomic relaxation induced by a laser noise. In the resonant case and electric field strengths small with respect to the atomic unit electric strength, the present nonperturbative results are in agreement with those obtained within the rotating wave approximation of Eberly et al. and Wodkiewicz et al., for an atom modelled by a two–level system. We discuss some examples which demonstrate the destruction of atomic coherence by the noise, the regime of relaxation to equilibrium state and the optical analogue of motional narrowing. We also give new results for two–level and three–level systems, and for a strong laser field at exact resonance, in the case of phase, amplitude noises. The case of fluctuation due to collisions is also discussed. Our numerical results indicate that ionisation effects, in the presence of laser noise, can lead to important modifications of the populations for strong laser–atom interactions. The changes generated onto the ionisation rates by the noise are also investigated.

# 1. Introduction

It is actually recognised that nearly all types of laser–atom interactions can be strongly affected by laser noise. In fact, real atoms experience a fluctuating environment of many perturbing interactions and ideal lasers exist only in theoretical models, while the used laser sources are subjected to many types of fluctuations notably in phase, amplitude and frequency. Stochastic variations of the Hamiltonian can be due not only to the field but also to the jump–type transitions of the atom from one energy state to another, jumps which are characteristic of condensed phases and which modulate its interaction with the medium [1]. Therefore, we cannot realise, without taking into account the statistical properties of the laser radiation, an exact comparison between theoretical predictions and experimental results. Incorporation of such stochastic properties into the Liouville equation by fully microscopic treatment (relaxation times and bandwidths) [2] give modest results, so we use a theoretical models based on the Markov pre–Gaussian processes [2–5]. These processes, composed of N–independent two–state jump processes (random telegraphs), form a non–Gaussian stochastic models, also called Markov chains. Such a Markov chain offers a detailed discussion of the atomic response. Our choice of the Markov chain is based on the simplicity of this model and the remarkable property of convergence to a Gaussian stochastic process when $N \to +\infty$. We derive the proper master equation for the density operator driven by a noise described by two–state stochastic telegraph process. This fundamental equation first introduced in quantum optic by Burshtein [2–4], contains all information concerning both the atomic transitions dynamic and the stochastic evolution of laser field fluctuations.

Several works have reported on the action of random process on a two–level system [1-4,6,7], particularly the evolution populations. In the present article we want to investigate the atomic response to a noise laser, i.e., to describe the relaxation of the atomic level populations $\rho_{nn}$. It should be noted that the kinetic of the phase relaxation contains information on the width and shape of the spectral lines [1]. We also extend our study to a three–level system and we provide the ionisation influence for strong

laser–atom interactions, when the laser field parameters are subjected to random process.

We are concerned here with an important theme of contemporary research, namely the interplay between quantum coherence and external noise. The destruction of quantum coherence by noise is central to many fields of physics and is reflected in the large number of papers recently published on this subject [8-12].

In this paper we are interested by a strong laser field whose frequency is resonant; time dependent perturbation theory and rotating wave approximation (RWA) [13] are therefore not adequate to resolve the interaction process. Indeed, the perturbation method is justified when the electric field strength $F_0$ remains much smaller than the atomic unit of field strength, namely $F_0 \ll 5.10^9$ V cm$^{-1}$ and when the laser photon energy is not tuned close to an atomic transition energy. The second limitation is resolved by the RWA, however there is the Bloch–Sierget shift [13] which deteriorates the efficiency of this approximation, especially when the laser intensity increases. We explore the limits of validity of rotating wave approximation for multi-level system and/or for strong laser field. We have therefore developed a nonperturbative treatment of the laser–atom interactions, based on the Floquet theory [14-18]. This nonperturbative method allows to transform the time dependent Liouville or Schrödinger equation to a stationary problem of eigenvalues. This problem can be solved without restrictions on the laser parameters, i.e., its solution requires at most a finite matrix diagonalization. It is another advantage of Floquet method in the context of laser–atom interactions.

The paper is structured as follows. In section 2, we present a detailed description of the theoretical formalism, which is valid for the general case of multi-level atoms. By considering the case of noise strong laser–atom interactions, for which detailed Floquet calculations are feasible to obtain the density matrix elements, which are solutions of the master equation. The account of Floquet theory given here is rather brief, since the theory has been discussed at length in the recent literature (see e. g. refs. [14–18]). Numerical results concerning the model of two–level and three–level systems are presented and discussed in section 3, where the phase and amplitude noises are described by random telegraph and Markov chain. The case of fluctuations due to collisions is analysed. We also discuss the noise effect on the ionisation process. At the end a summary of our results is given.

## 2. General formalism

We consider a model of multi–level atom excited by a classical purely monochromatic laser field described by an electric field linearly polarised,

$$\mathbf{F}_0(t) = \mathbf{F}_0 \cos(\omega t + \phi(t)). \qquad (1)$$

Here $F_0$ is the electric field amplitude (possibly fluctuating in magnitude) and $\phi(t)$ is the instantaneous phase of the laser ( fluctuating around the zero value ). The laser parameters are affected by a stochastic process of jumps. We represent these fluctuations by a Markovian Pre–Gaussian models. Thereafter, we denote the random telegraph and the Markov chain respectively [3]; by the processes [ $x(t) = \pm a$ ] and [ $X(t) = \{-Na, -(N-2)a, ..., (N-2)a, \text{and } Na\}$ ], with a is the amount of the jump assigned to the stochastic signal and N the number of random telegraph signals [3].

In order to incorporate a laser phase or amplitude noise in equations, treating the laser–atom interactions, we have therefore describe the system states considered in terms of the density matrix elements. The time dependent behaviour of the density operator $\rho$ is given by the Liouville equation [18]

$$i\hbar \frac{\partial \rho}{\partial t} = [H(s), \rho] - \frac{i\hbar}{2}[\Gamma, \rho]_+ + i\hbar \Lambda, \qquad (2)$$

where $\Gamma$ and $\Lambda$ are two diagonal matrix corresponding to the spontaneous emission process and defined by [18]

$$\Gamma_{nn} = \sum_{n'<n} \gamma_{nn'} \quad \text{and} \quad \Lambda_{nn} = \sum_{n'>n} \gamma_{nn'} \rho_{n'n'}(t), \qquad (3)$$

where $\gamma_{nn'}$ is the radiative decay rate. $H(s)$ is the non–relativistic periodical Hamiltonian in the presence of the laser noisy and s labels any possible state of the random process. In the case of the laser amplitude fluctuations, $H(s)$ may be written as

$$H(s) = H_0 + e\, \mathbf{F}_0(1 + x(t))\cdot\mathbf{R} \cos(\omega t + \phi(t)) \qquad (4)$$

and, in the case of the laser phase fluctuations, it takes the following form

$$H(s) = H_0 + e\, \mathbf{F}_0\cdot\mathbf{R} \cos(\omega t + \phi(t) + x(t)), \qquad (5)$$

where $H_0$ is the nonperturbative Hamiltonian, $\omega$ the frequency of laser field, e is the electric charge and $\mathbf{R}$ the dipole operator.

In the stationary eigenstates $\{|n\rangle\}$ of the unperturbed system Hamiltonian, the equation (2) could be written in the following system of coupled differential equations for the time dependent matrix elements density $\rho_{nn'}(t)$ as

$$\dot{\rho}_{nn'} = -i\omega_{nn'}\rho_{nn'} - \frac{i}{\hbar}\sum_k \left[ \rho_{kn'} M_{nk}(s) - \rho_{nk} M_{kn'}(s) - \frac{i\hbar}{2}(\Gamma_{nk}\rho_{kn'} + \rho_{nk}\Gamma_{kn'}) \right] + \delta_{nn'}\Lambda_{nn} \qquad (6)$$

where $\omega_{nn'} = \dfrac{E_n - E_{n'}}{\hbar}$ is the Bohr frequency associated with the n → n' transition, $M_{nk}(s)$ the dipole coupling matrix elements in presence of fluctuations, corresponding to stochastic state s of the random process and $\delta_{nn'}$ is the Kronecker symbol.

Since the process that we are considering here is Markovian, the conditioned probability density function associated with it, namely $p(s,t|s_0,t_0)$, is shown to satisfy the following Chapman–Kolmogorov equation [2–4,19,20]

$$\frac{\partial}{\partial t}p(s,t|s_0,t) = -\frac{N}{T}p(s,t|s_0,t_0) + \frac{N}{T}\left[\frac{N+s}{2N}p(s-2,t|s_0,t_0) + \frac{N-s}{2N}p(s+2,t|s_0,t_0)\right], \qquad (7)$$

here $s_0$ is the initial state of Markov chain at the time $t_0$.

In the simple case of random telegraph signal, the above equation reduces to

$$\frac{\partial}{\partial t}p(s,t|s_0,t_0) = -\frac{1}{T}p(s,t|s_0,t_0) + \frac{1}{T}p(-s,t|s_0,t_0). \qquad (8)$$

In the compact form, the equation (8) write as

$$\frac{d\mathbf{P}_s}{dt} = \mathbf{W}_s^{s'} \mathbf{P}_{s'}, \tag{9}$$

with $\mathbf{W}_s^{s'} = \frac{1}{T}\begin{bmatrix} -1 & 1 \\ 1 & -1 \end{bmatrix}$ is the relaxation matrix composed by the frequencies of telegraph jumps process, where s and s' are two different states of random telegraph (s and s' = 1,2), corresponding to the telegraph signal amplitude ($\pm a$). T denote the dwell time (i.e., the mean time between interruptions) for the telegraph.

The main difficulty of typical problems lies in the correct averaging of the matrix density over all realisations of noise. In fact, what is wanted is $\langle \rho_{nn'} \rangle$, that is, the solution to the equation (6) averaged over the ensemble of jumps of the implicit telegraph x(t). To obtain $\langle \rho_{nn'} \rangle$ one proceeds indirectly, defining a marginal average $\rho_{nn',s}$ (t), by the equation

$$\langle \rho_{nn'} \rangle = \sum_s g(s)\, \rho_{nn',s}, \tag{10}$$

where g(s) is the initial probability distribution of the random process and $\rho_{nn',s}$ (t) the average value of $\rho_{nn'}$(t) under the condition that x(t) is fixed at the value s at time t. Combining the Chapman–Kolmogorov equation (9) for the probability density function and the dynamic equation (6) for the statistical operator $\rho(t)$, a master equation can be derived for the so–called marginal averages $\rho_{nn',s}$ (t) [2,3]. It reads

$$\frac{\partial \rho_{nn',s}}{\partial t} = -\frac{i}{\hbar}\sum_k \left[ \rho_{kn',s} M_{nk}(s) - \rho_{nk,s} M_{kn'}(s) - \frac{i\hbar}{2}(\Gamma_{nk}\rho_{kn',s} + \rho_{nk,s}\Gamma_{kn'}) \right] \\ -i\omega_{nn'}\rho_{nn',s} + \delta_{nn'}\Lambda_{nn,s} + \mathbf{W}_s^{s'}\rho_{nn',s'} \tag{11}$$

with

$$\Lambda_{nn,s} = \sum_{n'>n} \gamma_{nn'}\, \rho_{n'n',s}(t). \tag{12}$$

The equation (11) exhibits a system of differential equations with periodical coefficients. This system constitutes a fundamental equation for any statistical study of the interaction processes in the presence of fluctuations. By using the usual Floquet technique [14–18], we can seek the solution of the form

$$\rho_{nn',s}(t) = e^{-i\varepsilon t/\hbar} \sum_{M=-\infty}^{+\infty} e^{-iM\omega t} \; C_{nn',s}^{M}. \tag{13}$$

The Floquet coefficients $C_{nn',s}^{M}$ and the pseudo–energies $\varepsilon$ can be found by solving numerically the eigenvalues problem

$$\begin{aligned}&\frac{1}{\hbar}\sum_{k}\left[C_{kn',s}^{M+1} M_{nk}^{+1}(s) + C_{kn',s}^{M-1} M_{nk}^{-1}(s) - C_{nk,s}^{M+1} M_{kn'}^{+1}(s) - C_{nk,s}^{M-1} M_{kn'}^{-1}(s)\right] \\ &- \frac{i}{2}\sum_{k}\left[\Gamma_{nk} C_{kn',s}^{M} + C_{nk,s}^{M} \Gamma_{kn'}\right] + (\omega_{nn'} - M\hbar\omega) C_{nn',s}^{M} + i\delta_{nn'} \sum_{n'>n} \gamma_{nn'} C_{n'n',s}^{M} \\ &+ iW_{S}^{S'} C_{nn',s'}^{M} = \varepsilon \; C_{nn',s}^{M} \end{aligned} \tag{14}$$

where the dipole coupling matrix elements $M_{nk}^{-1}(s)$ and $M_{nk}^{+1}(s)$ are defined by

$$M_{nk}(s) = M_{nk}^{-1}(s) \; e^{-i\omega t} + M_{nk}^{+1}(s) \; e^{+i\omega t} \tag{15}$$

In matrix representation, the system of equations (14) could be written in the following compact form, as

$$\mathbf{G}\,\psi = \varepsilon\,\psi \tag{16}$$

where $\mathbf{G}$ is the infinite tridiagonal block Floquet matrix, $\varepsilon$ and $\psi$ the eigenvalues and the eigenvectors of $\mathbf{G}$. In numerical calculations, the series expansion in function of M (expression (13)) is truncated to a finite $M_{max}$ number of terms which permits the convergence of eigenvalues and eigenvectors of $\mathbf{G}$. In the absence of noise, the density operator satisfies a 9×9 matrix differential equation with constant coefficients ( case of three–level system ). The master equation allows stochastic telegraph type noise to be added at the sole expense to enlarge the matrix dimensionality from 9×9 to $9n_s \times 9n_s$; $n_s$ represents here the two states of random telegraph signal. When we consider the Floquet solution based on the development of Fourier, we enlarged also the matrix dimensionality from $9n_s \times 9n_s$ to $9n_s(2M_{max}+1) \times 9n_s(2M_{max}+1)$, with $(2M_{max}+1)$ is the number of Floquet matrix blocs. We remark that dimensionality of $\mathbf{G}$ becomes very large [dim **G** = $n^2 \times n_s \times (2M_{max}+1)$, with n is the number of atomic states] and depends of $M_{max}$. This latter is the truncated number of Fourier development permitting the convergence of eigenvalues and eigenvectors of Floquet matrix. It is important to remark the following fact, the Floquet pseudo–energies $\varepsilon_j$ are only defined modulo the photon energy, so the Floquet quasienergy are not unique. We extract the pseudo–energies $\varepsilon_j$, which are physically independents, from the Floquet matrix eigenvalues.

The Floquet solution of the system (14) is a linear combination of all solutions corresponding to different pseudo–energies $\varepsilon_j$, and their eigenvectors $C_{nn',s}^{M,\varepsilon_j}$, we set

$$\rho_{nn',s}(t) = \sum_{j=1}^{n^2 \times n_s} \alpha_j \, e^{-i\varepsilon_j t/\hbar} \sum_{M=-M_{max}}^{+M_{max}} e^{-iM\omega t} \, C_{nn',s}^{M,\varepsilon_j} \qquad (17)$$

where $\alpha_j$ are the constants of linear combination, determined by the initial conditions [$\rho_{11}(t=0) = 1$ and $\rho_{22}(t=0) = \rho_{33}(t=0) = 0$]. On averaging over all realisations of the random signal, we obtain the final expression of matrix density elements $\rho_{nn'}(t)$ as

$$\rho_{nn'}(t) = \sum_{s=1}^{2} g(s) \sum_{j=1}^{n^2 \times n_s} \alpha_j \, e^{-i\varepsilon_j t/\hbar} \sum_{M=-M_{max}}^{+M_{max}} e^{-iM\omega t} \, C_{nn',s}^{M,\varepsilon_j} \qquad (18)$$

We remark that the Floquet method permits us to evaluate exactly atomic response functions in finite terms, and we can then examine the influence of laser fluctuations on these functions.

# 3. Results and discussion

By describing external sources of noise by Pre–Gaussian Markov chains composed of N–independent jump processes, we treated two different sources of pre–Gaussian noise (phase and amplitude) by using a simple general master equation soluble in finite terms. Our application is carried out in two cases. In first, the model of two–level atom on resonance is considered and it results compared with those obtained by Eberly et al. and Wodkiewicz et al. [2,3]. We also provide some results for atomic response, when RWA approximation is not justified for strong laser fields. In second, the behaviour of three–level atom response to laser noisy is investigated. We choose the inverse Rabi frequency $\Omega$ as time unit, (where $\Omega = \mathbf{d}_{nk}.\mathbf{F}_0$, with $\mathbf{d}_{nk}$ is the dipole matrix element between levels $|n\rangle$ and $|k\rangle$), in the aim to analyse the obtained results in terms of the noise strength.

In the absence of any noise source ( electric laser field is purely coherent) and if we neglect the other relaxation rates ( spontaneous decay and ionisation rate), the behaviour of the populations of two-level and three–level systems is illustrated in Figure 1., which shows a superposition of several undamped oscillation modes ( Rabi oscillations). The system remains indefinitely in this oscillatory state.

### 3. 1. Two–level atom in presence of random telegraph laser noise.

As already noted by several authors [2–4], the behaviour of a two–level atom to pre–Gaussian noise in strong laser–atom interactions depends critically on a special telegraph noise. The agreement between the two methods (the numerical calculation within the framework of rotating wave approximation and the present nonperturbative method of calculation based on the Floquet theory) is good at resonant excitation and when the electric field strength $F_0$ remains much smaller than the atomic unit of field strength $5.10^9$ V/cm. This behaviour is illustrated in Figure 2.(A), where we show an example of phase random telegraph influence on atomic level populations corresponding to the phase jumps $a = 0.4\pi$, the electric field strength $F_0 = 10^6$ V/cm for the phase switching rates ($\Omega T = 0.1$, 1 and 10). Figure 2.(B) displays the atomic level populations resonantly excited by random telegraph phase noise, corresponding to the phase jumps a = 0.4π, the electric field strength $F_0$ = 5. $10^8$ V/cm and from phase fluctuations for three different rates switching. The regime of weak damping is observed for ($\Omega T$ = 10). The atomic populations exhibit a motional narrowing regime for small Rabi frequency compared to the frequency noise ($\Omega T$ = 0.1), in this case, the telegraph jumps are too fast that the atomic system can feel only the mean value of fluctuations. For ($\Omega T$ = 1), we note a partial destruction of the atomic coherence by the phase noise and the kinetic of relaxation is rapid (strong damping). The two populations converge to a steady-state of value 1/2, in indication that the phase jump relaxation is purely ''transverse''[2]. It is interesting to note the presence of an irregular behaviour on the oscillations of the two populations for a strong laser field; in fact, we observe small oscillations which come to superpose to the Rabi oscillations, their amplitude is weak and disappears when the electric field strength $F_0$ becomes small with respect to the atomic unit of field strength. These little oscillations represent the fast variable phases $e^{\pm i (\omega+\omega_{nn'})t}$ which are safely neglected by the authors [2–4], by using the rotating wave approximation. Figure 3 shows an example of amplitude telegraph influence on atomic level populations. Taking a = 0.1, three different amplitude switching rates ($\Omega T$ = 0.1,1,100) and the electric field strength is $F_0$ = 5. $10^8$ V/cm. We remark a very weak damping at ($\Omega T$ = 0.1,1) and a beats phenomena at ($\Omega T$ = 100). In order to lead the system to the relaxation process, we must use a large number of Rabi periods than in the case of phase fluctuations. The main difference observed between the random telegraph phase and amplitude modulation arises from the choice of the laser stochastic fluctuations.

As already mentioned in the introduction, under collision effects, the transition frequency $\omega_{21}$ can also fluctuates around its fixed value $\omega_{21}$. The simplest model of such interruption collisions [4,21] assumes that the atomic transition frequency $\omega_{21}$ should be replaced by $\omega_{21}(t) = \omega_{21} + x(t)$. Figure 4 displays , the influence of such collisional noise on the atomic response, where we take the electric field strength $F_0$ = 5. $10^8$ V/cm, the jump parameter a = 0.1 and three different frequency switching rates ($\Omega T$ = 1, 10 and 100). We remark damped quasiperiodic oscillations. The case of ($\Omega T$ = 10) corresponds to strong damping without any convergence to a steady state. While the relaxation to a equilibrium state of value 1/2 is clear for a switching rate ($\Omega T$ = 1). The damping becomes weak for ($\Omega T = 100$) and two independent beats phenomena are observed. The complicated time evolution of populations is a result of Rabi oscillations interference.

### 3. 2. Two–level atom in presence of Markov chain laser noise.

We generalise the case of random telegraph to a Markov chain, which treats the phase and amplitude fluctuations of strong laser field interacting resonantly with a two level system. Figure 5. shows the atomic response for a Markov chain composed of three and seven phase telegraphs, with the phase jump parameter is $a = 0.4\pi$, the electric field strength is $F_0 = 10^8$ V/cm and ($\Omega T = 1, 100, 1000$). For ($\Omega T = 1$), we remark that the phase noise entirely eliminates the atomic coherence. The Rabi oscillations have been completely destroyed and the relaxation process is rapid, while for the cases ($\Omega T = 100$ and $\Omega T = 1000$) the coherence effects are restored even though the field is fluctuating. The damping strength decreases progressively for ($\Omega T = 100$) and ($\Omega T = 1000$). Similar behaviour is also observed when we increase the number of phase telegraphs (seven telegraphs), but the damping becomes more intense and we clearly observe the convergence to a Gaussian limit.

Figure 6. shows the same situation, as Figure. 5, but for amplitude noise with the jump parameter $a = 0.1$. New beats phenomena which appear at large switching rates ($\Omega T = 100, 1000$). We remark that the degree of Gaussian character and the kinetic of convergence to a stationary state increase with the number N of random telegraphs. This behaviour justifies the pre–Gaussian property of Markov chains.

### 3. 3. Three–level system in presence of random telegraph laser noise.

Multilevel systems show a variety of interesting optical effects with laser fields [22-23]. It would be interesting to study how the incoherence sources affect the population evolution of multilevel atoms. For the sake of numerical simplicity (the Floquet matrix dimentionality largely increases from two-level to multi-level), we only extend our application to a three-level atom driven by stochastic strong laser field at resonance. Our model of three-level atom is a *ladder* system which contains three discrete bound states and a continuum. It may be considered as a generalisation of the so-called ''extended two–level'' model proposed by Yeh and Eberly [24], where they have assumed that the bound–continuum dipole moments are weakly energy dependent so that it is well justified to use the adiabatic following elimination of the continuum degrees of freedom [24]. In absence of ionisation, no loss out of the system occurs, but damping is supplied by spontaneous decay and laser noisy within the three-level system, so that non zero populations are maintained in a steady-state for large times. Figure 7 clearly illustrates this behaviour in the case of one phase telegraph, with a = $0.4\pi$, three different phase switching rates ($\Omega T$= 0.1, 1 and 10) and the electric field strength is $F_0$ = 5. $10^8$ V/cm. The relaxation process is observed for ($\Omega T = 1$) with strong damping. The cases of ($\Omega T = 0.1$) and ($\Omega T = 10$) correspond to weak damping. The three populations $\rho_{11}$ (t), $\rho_{22}$ (t) and $\rho_{33}$ (t) converge to a stationary state of value 1/3.

Figure 8. shows the case of atomic response for amplitude noise with a = 0.1, $F_0$ = 5. $10^8$ V/cm and ($\Omega T$= 1, 10, 100). For ($\Omega T = 1$) we have a weak damping and a strong damping for ($\Omega T = 10$), the populations $\rho_{11}$ (t) and $\rho_{33}$ (t) converge to a stationary state of value 3/8, while $\rho_{22}$ (t) converge to a stationary state of value 1/4 with rapid damping than for $\rho_{11}$ (t) and $\rho_{33}$ (t). Quantum beats appear for ($\Omega T = 100$) between $\rho_{11}$ (t) and $\rho_{33}$ (t), while $\rho_{22}$ (t) shows beats independently. It is apparent that important asymmetries between the atomic response in the case of phase and amplitude fluctuations should be expected. This difference is justified by the fact that in the case of amplitude fluctuations, the jump parameter a, assigned to stochastic process, appear in term of laser intensity $F_0.(1 \pm a)$, while in the case of phase noisy , the dependence occurs in term of ($e^{\pm i a}$).

**3.4 Ionisation effects on field–atom interactions in presence of laser noise.**

Since we have considered a strong laser field, we would have a large ionisation for all atoms. In order to take into account of the noise laser effects on the atomic populations and ionisation process, we will incorporate the responsible term of ionisation [$\tau_{EC} = -R_{nc}\delta_{nn'} -1/2(R_{nc} + R_{n'c})(1-\delta_{nn'})$, where $R_{nc}$ is the relaxation rate from the excited state $|n\rangle$ to the continuum $|c\rangle$] in the motion equations (2), (6), (11) and (14). Two ionisation probabilities are calculated respectively the instantaneous probability $P_I(t) = 1 - \eta(t)$ where $\eta(t) = \sum_{n=1}^{3} \rho_{nn}$, and the mean ionisation probability $P_{Im}(t) = 1 - e^{-R_I t}$ where $R_I$ and we represent the effect of different laser noises on the average ionisation probability $P_{Im}(t)$. Same behaviour can be observed for the ionisation rate $R_I$. Figure 9(a) displays the time evolution of the two-probabilities $P_I(t)$ and $P_{Im}(t)$ in absence of noise. Figure 9(b) illustrates the influence of one amplitude telegraph on the average ionisation probability $P_{Im}(t)$. The same parameters as Figure 6 are used. The results depend on the fluctuations time scale compared to the other characteristic time scales of the problem. The minimum ionisation rate is obtained for ($\Omega T = 10$). Figure 9(c) shows a weak effect of phase noise on the ionisation probability and minimum ionisation rate is obtained for ($\Omega T = 1$).

Figure 10. shows the response of two–level (A column) and three–level (B column) systems in the case of laser phase noise described by a random telegraph. The same parameters are taken as Figure 2.(B column) and Figure 7. The novelty in this figure is the incorporation of ionisation process, which is represented by the instantaneous ionisation probability $P_I(t)$. A loss of population has been induced by ionisation, indeed, we remark that the populations oscillate in the same manner that in absence of ionisation effect, but there is a progressive decay to the zero probability, while the ionisation probability $P_I(t)$ increases in time. The same behaviour is observed in the case of laser amplitude noise and for a Markov chain.

# 4. Conclusion

In this paper, we have presented a general stochastic treatment to incoherence properties induced by laser fluctuations (phase and amplitude) and by collision effects (frequency). Our method is based on the nonperturbative Floquet theory with pre-Gaussian processes and collisional approach modelling the different sources of noise. A detailed discussion of the noise effects on the atomic response has been given by resolving a master equation. We have examined the behaviour of three–level systems at 'high' laser intensities where both the two-state approximation and the rotating wave approximation fail. Our treatment has given good results. In fact, we have not only reproduced the results of other authors [2,3] for two-level system, (when the ionisation effects are neglected) but also, we have established new interesting results concerning the little oscillations which appear on the populations $\rho_{nn}$. Our results show a destruction of the atomic coherence by the noise and a relaxation regime to equilibrium state. The damping rate or relaxation kinetic is related to the size order of fluctuations. We have also investigated the effect of noise on the ionisation rates. On the basis of these results, obtained for two–level and three–level systems, the Floquet approach is then useful in the nonperturbative treatment of the interaction processes in presence of external sources of noise, because any restriction on the laser parameters is imposed. The nonperturbative method is very convenient for analysing the effects of laser noise on multi–level and real atomic systems.

# Figure captions

**Figure 1.** Populations $\rho_{nn}$ versus time (in units of inverse Rabi frequency $\Omega$) for two–level (a) and three–level (b) atoms, resonantly excited by purely coherent laser (no fluctuations). The electric field strength is $F_0 = 5.\ 10^8$ V/cm and the emission spontaneous coefficients are $\gamma_{21} = \gamma_{32} = 1.9\ 10^{-5}$

**Figure 2.** The columns A and B represent the populations $\rho_{nn}$ versus time (in units of inverse Rabi frequency $\Omega$) for two–level atom resonantly excited by random telegraph phase noise, successive frames are for different values of phase switching rates $\Omega T = 0.1$, 1 and 10. The phase jump parameter is $a = 0.4\pi$, the emission spontaneous coefficients are $\gamma_{21} = \gamma_{32} = 1.9\ 10^{-5}$ and the electric field strengths are $F_0 = 10^6$ V/cm and $F_0 = 5\ 10^8$ V/cm, respectively, for the plots of the columns A and B.

**Figure 3.** Populations $\rho_{nn}$ versus time (in units of inverse Rabi frequency $\Omega$) for two–level atom resonantly excited by random telegraph amplitude noise, successive frames are for different values of amplitude switching rates $\Omega T = 0.1$, 1 and 100. The amplitude jump parameter is $a = 0.1$, the electric field strength is $F_0 = 5.10^8$ V/cm and the emission spontaneous coefficients are $\gamma_{21} = \gamma_{32} = 1.9\ 10^{-5}$.

**Figure 4.** Populations $\rho_{nn}$ versus time (in units of inverse Rabi frequency $\Omega$) for two–level atom resonantly excited by random telegraph frequency noise, successive frames are for different values of switching rates $\Omega T = 1$, 10 and 100. The jump parameter is $a = 0.1$, the electric field strength is $F_0 = 5.10^8$ V/cm and the emission spontaneous coefficients are $\gamma_{21} = \gamma_{32} = 1.9\ 10^{-5}$.

**Figure 5.** Populations $\rho_{nn}$ versus time (in units of inverse Rabi frequency $\Omega$) for two–level atom resonantly excited by Markov chain phase noise, composed of three

phase random telegraphs (N = 3) and an extension to a Markov chain of seven phase random telegraphs (N = 7). Successive frames are for different values of phase switching rates $\Omega T = 1$, $10^2$ and $10^3$. The electric field strength is $F_0 = 10^8$ V/cm, the phase jump parameter is $a = 0.4\pi$ and the radiative decay rates are $\gamma_{21} = \gamma_{32} = 1.9\ 10^{-5}$

**Figure 7.** Populations $\rho_{nn}$ versus time (in units of inverse Rabi frequency $\Omega$) for threelevel atom resonantly excited by a random telegraph phase noise, with phase noise $a = 0.4\pi$. Successive frames are for different values of switching rates $\Omega T = 0.1$, 1 and 10. The electric field strength is $F_0 = 5.10^8$ V/cm, the phase jump parameter is $a = 0.4\pi$ and the radiative decay rates are $\gamma_{21} = \gamma_{32} = 1.9\ 10^{-5}$.

**Figure 8.** Same as Figure **7.**, but for amplitude fluctuations and the jump parameter is $a = 0.1$.

**Figure 9.** Ionisation probability versus time (in units of inverse Rabi frequency $\Omega$). The electric field strength is $F_0 = 10^8$ V/cm, the radiative decay rates are $\gamma_{21} = \gamma_{32} = 1.9\ 10^{-5}$ and The relaxation rates from bound states to the continuum are $R_{2c} = R_{3c} = \Omega/10$.

(a) Solid line: mean ionisation probability $P_{Im}(t)$. Dotted line: instantaneous ionisation probability $P_I(t)$. Solid and dotted lines are in absence of noise.

(b) Solid line: mean ionisation probability in absence of noise. Dashed line: in presence of amplitude noise ($a = 0.1$) with $\Omega T = 1$. Dotted line: in presence of amplitude noise with $\Omega T = 10$.

(c) Same as (b), but for phase noise ($a = 0.4\pi$).

**Figure 10.** The columns A and B are, respectively, same as Figure 2.(B) and Figure **7.**, but take into account of the ionisation process represented by the probability $P_I(t)$. The relaxation rates from bound states to the continuum are $R_{2c} = R_{3c} = \Omega/10$.

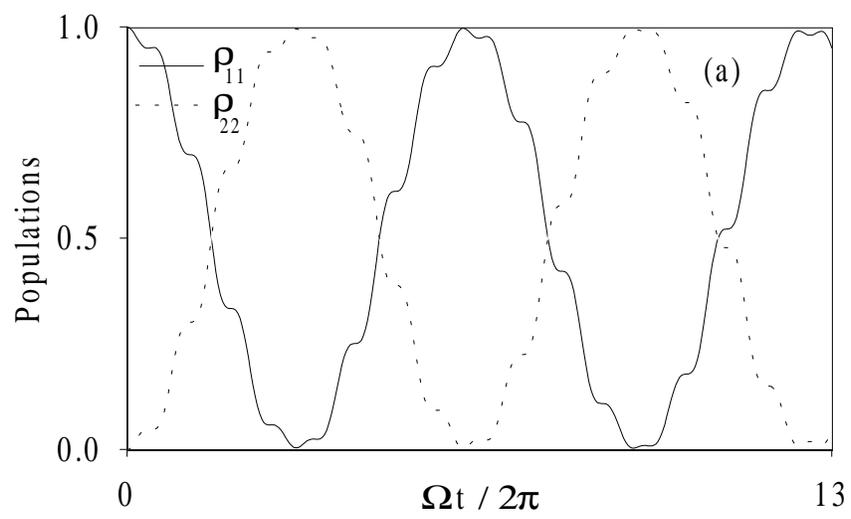

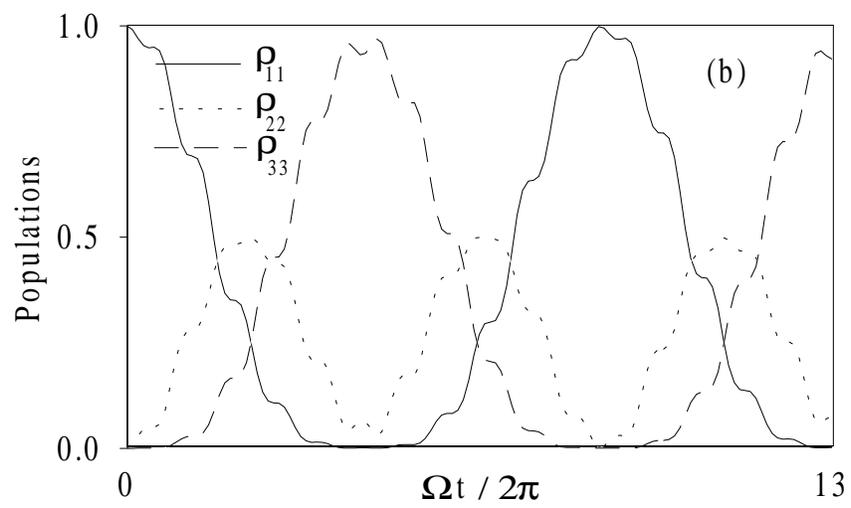

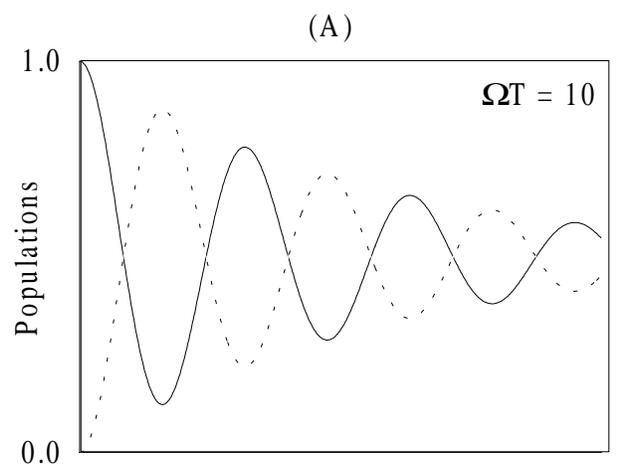
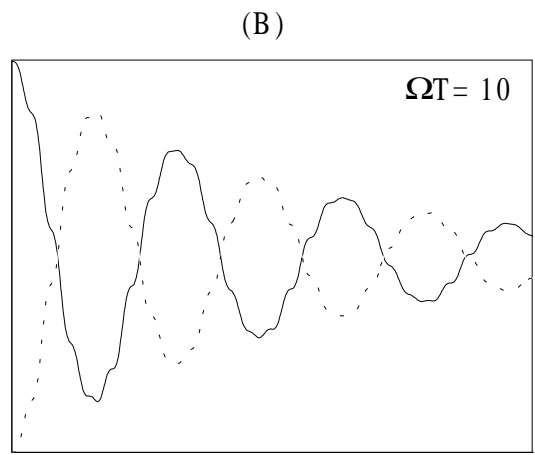
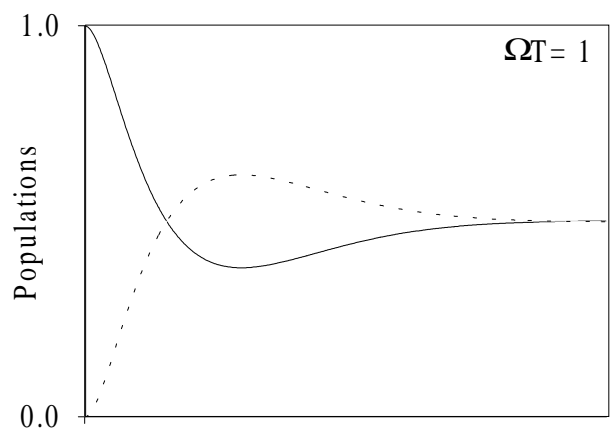
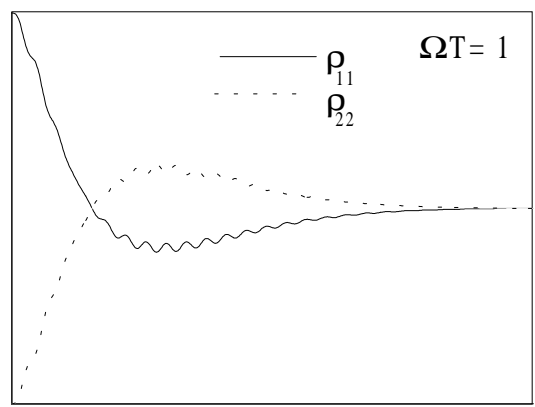
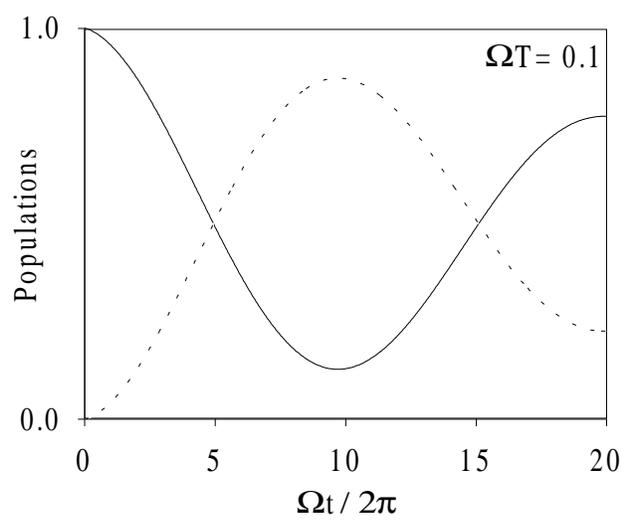
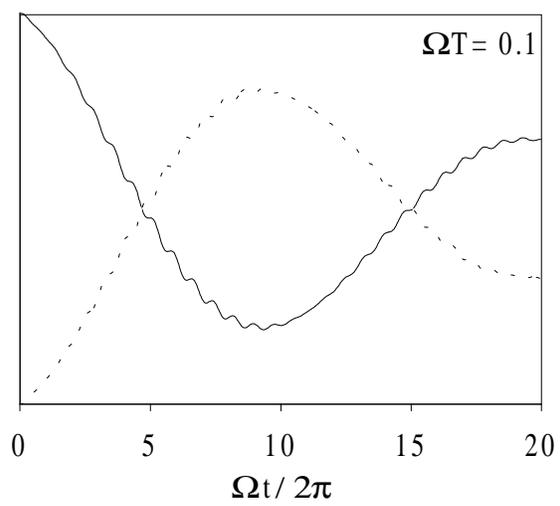

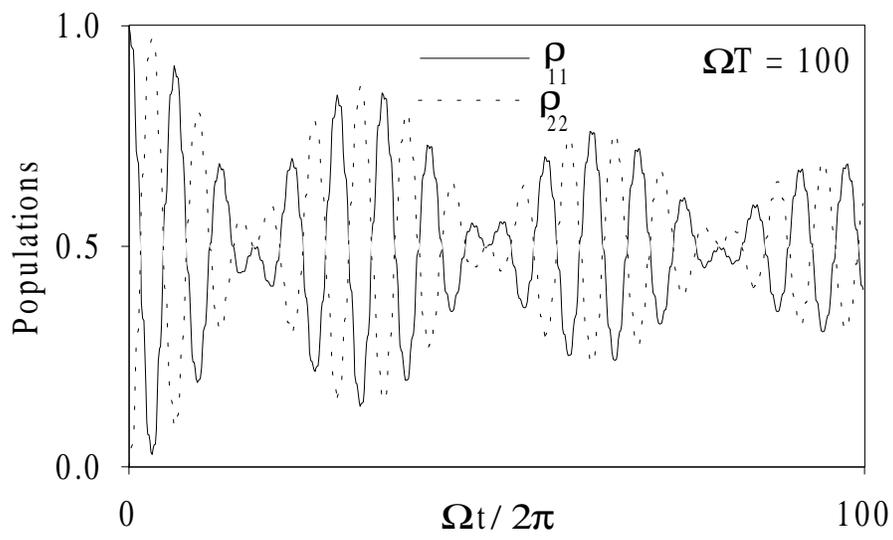

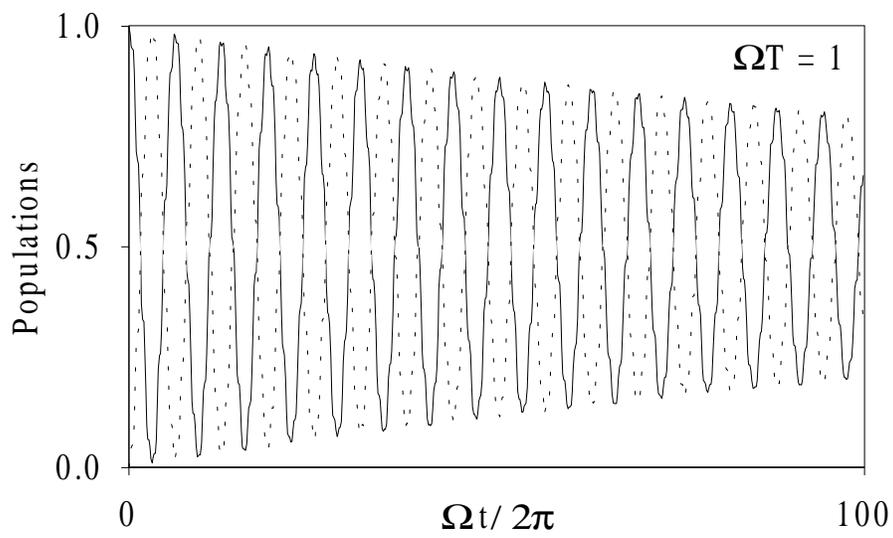

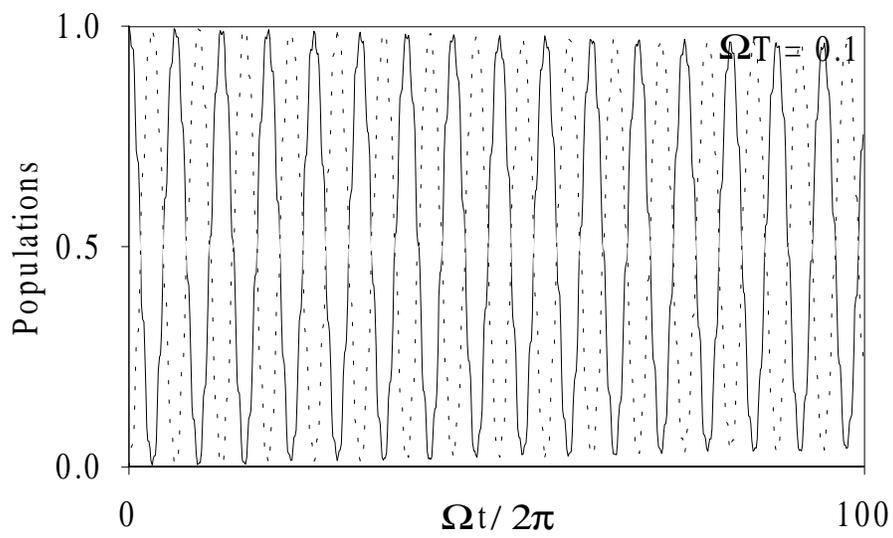

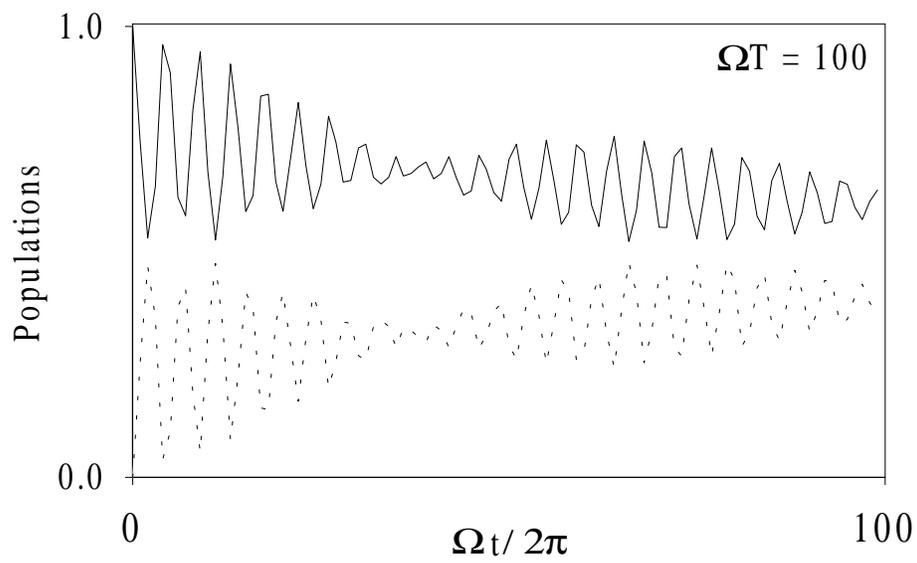
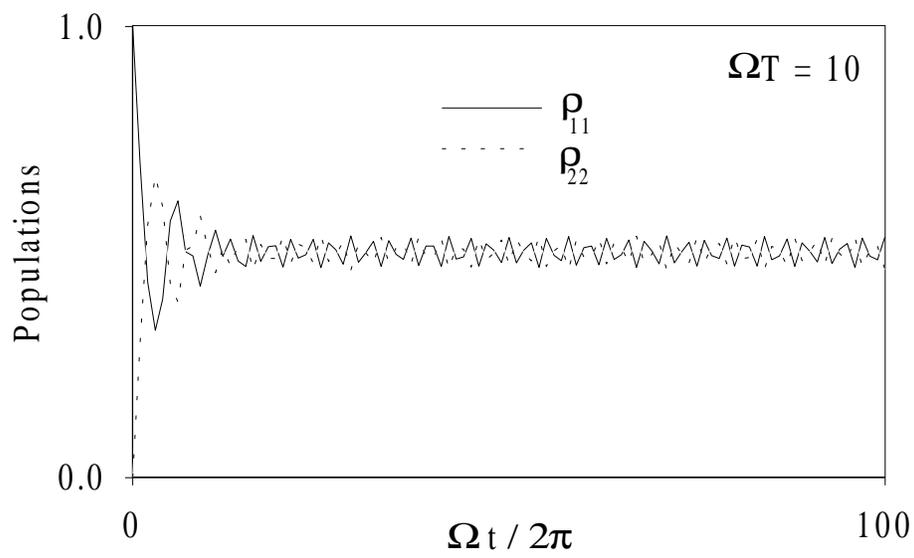
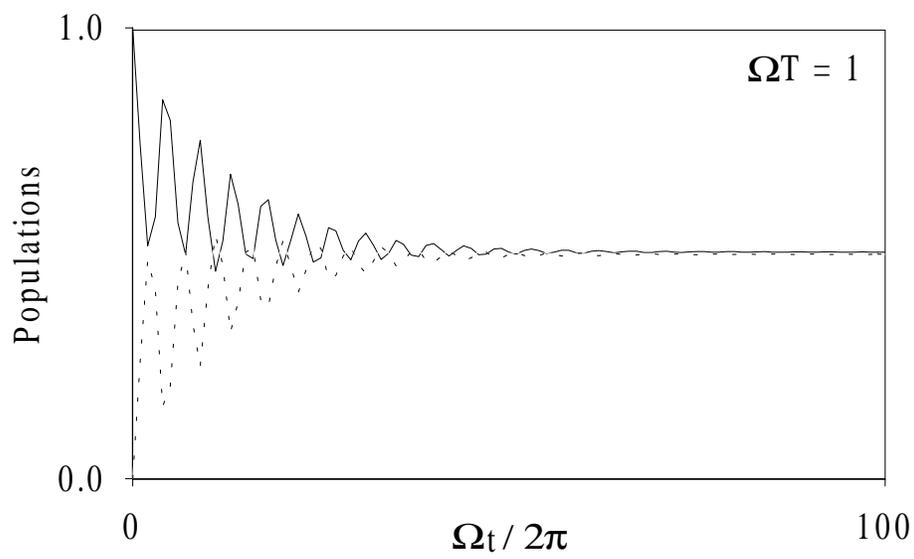

| (N = 3) | (N = 7) |
|---|---|
| 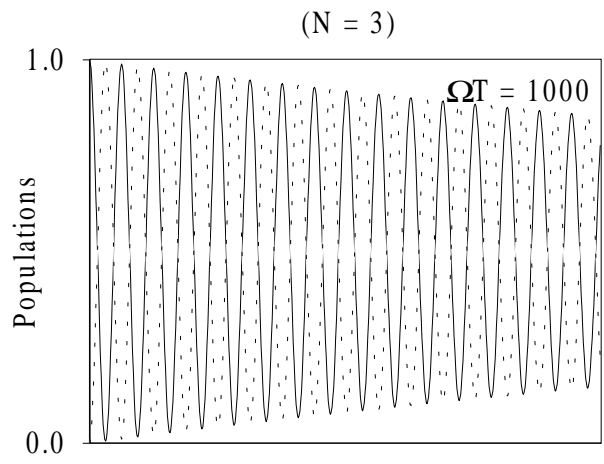 | 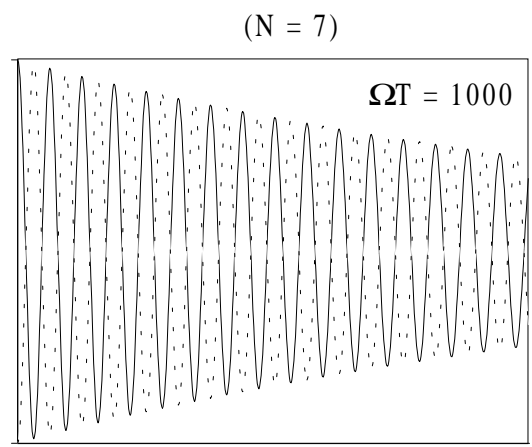 |
| 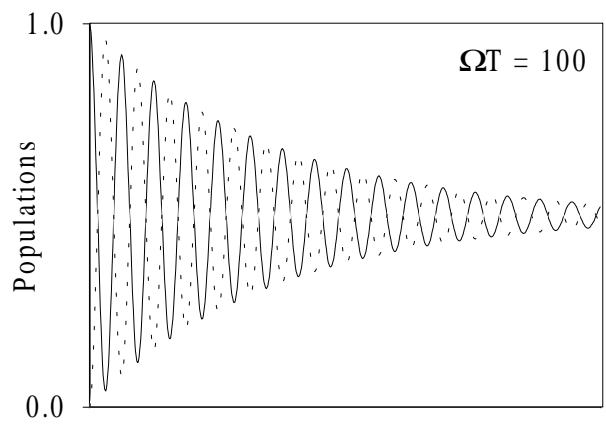 | 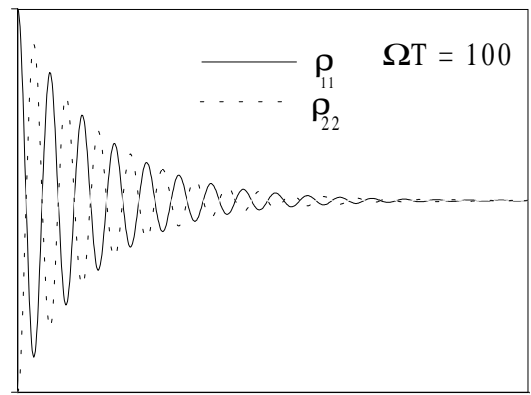 |
| 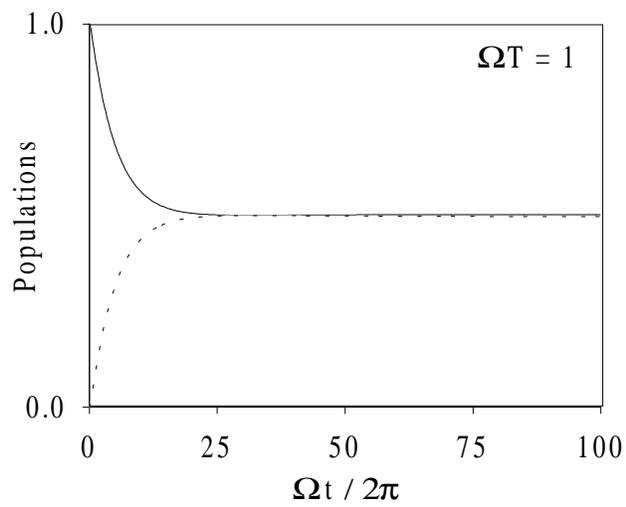 | 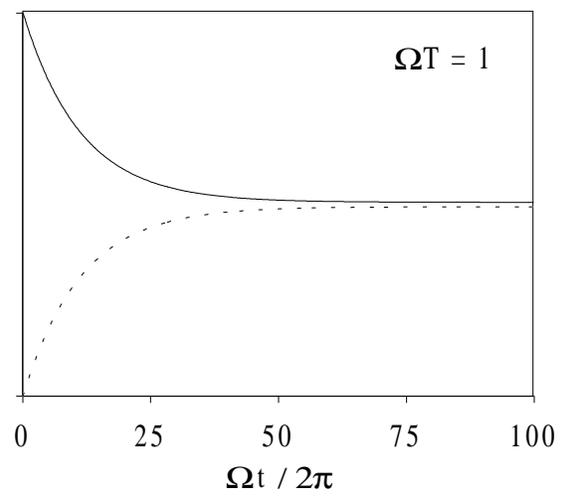 |

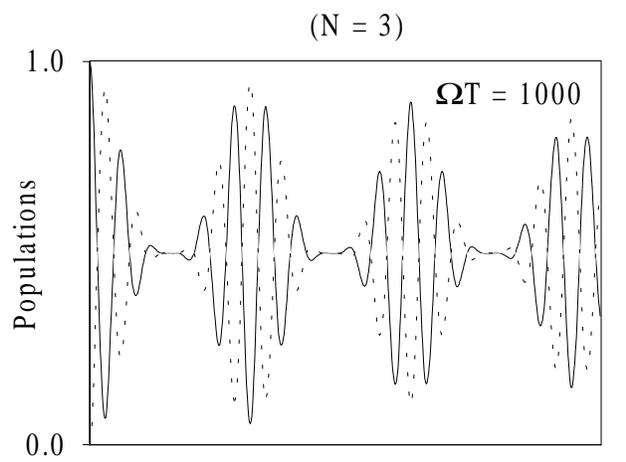
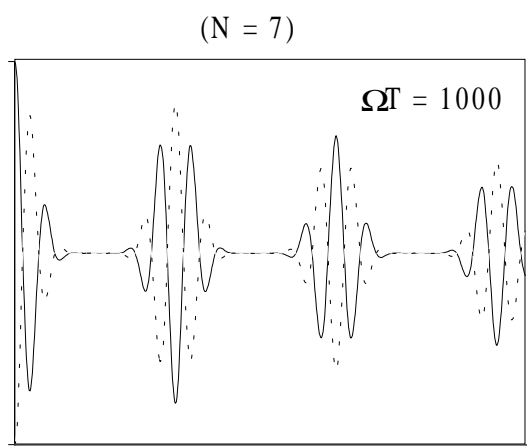
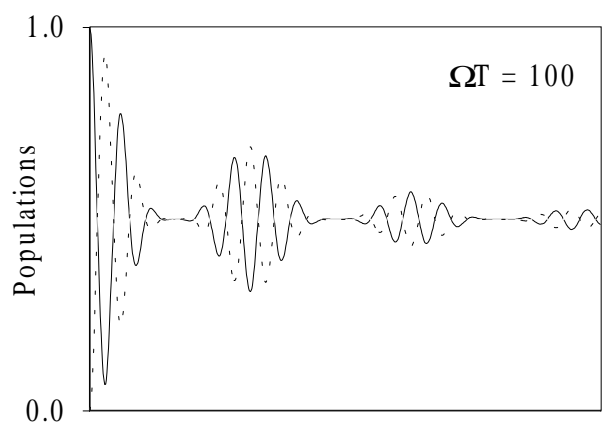
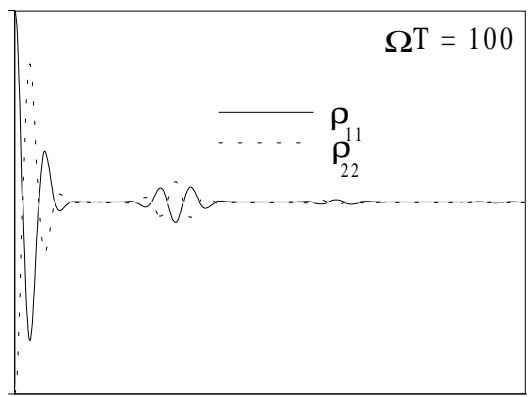
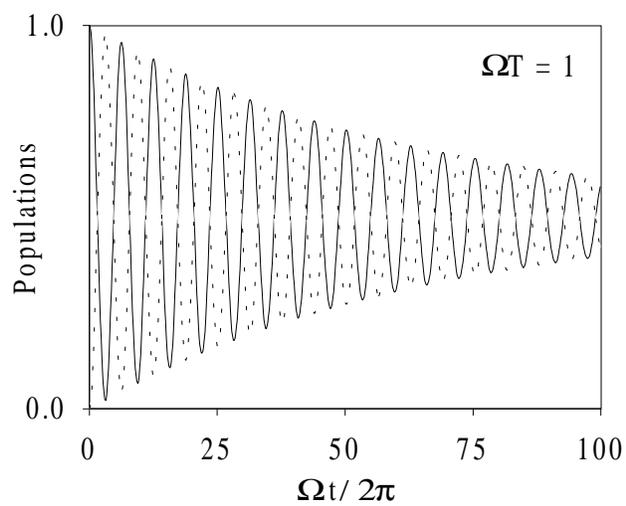
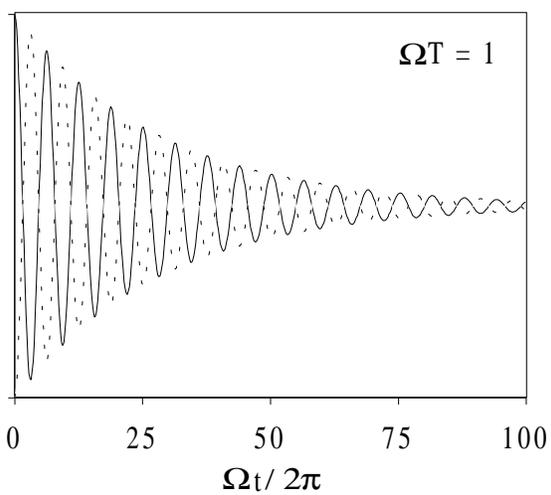

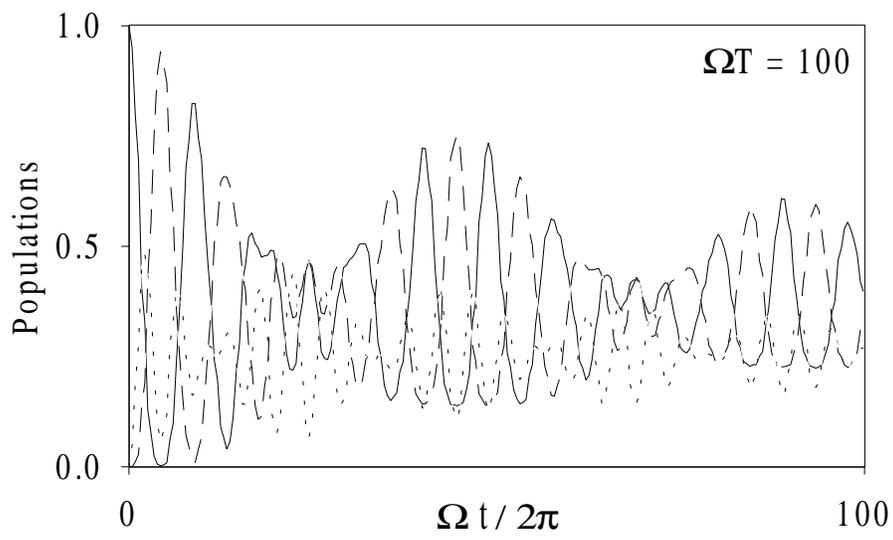

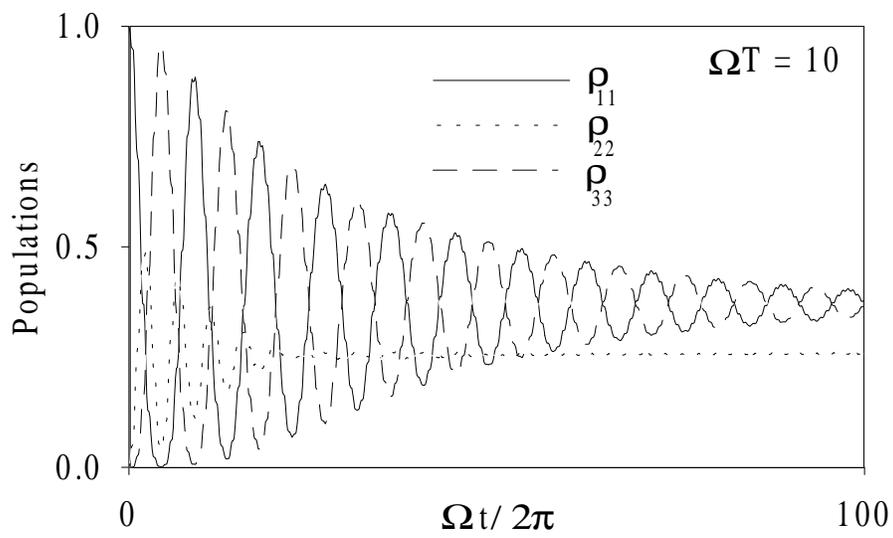

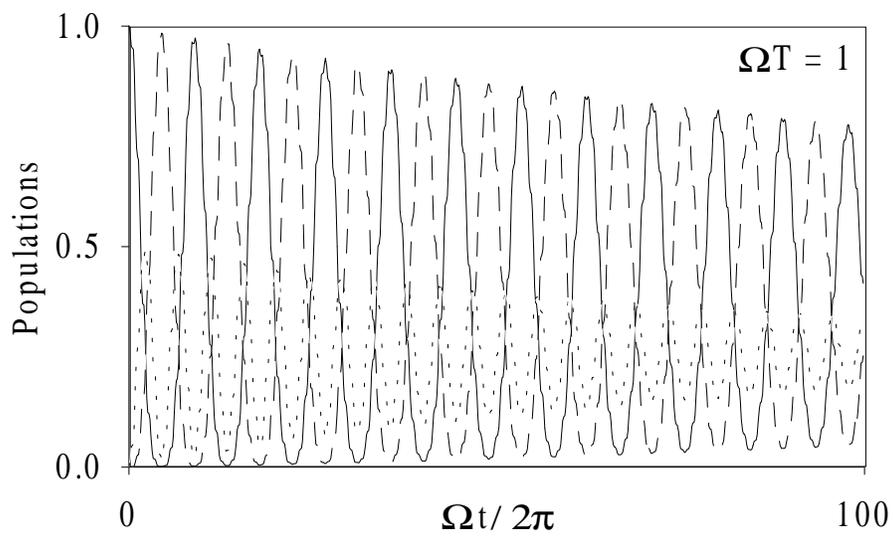

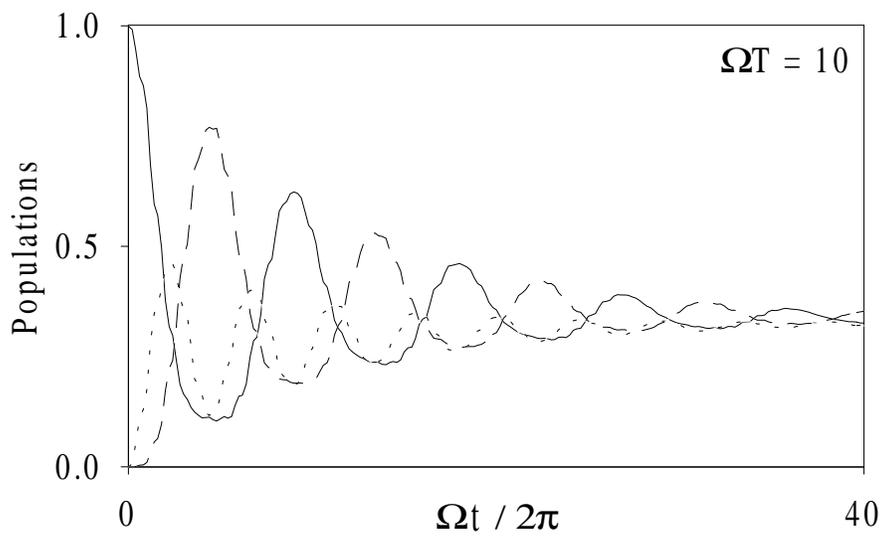

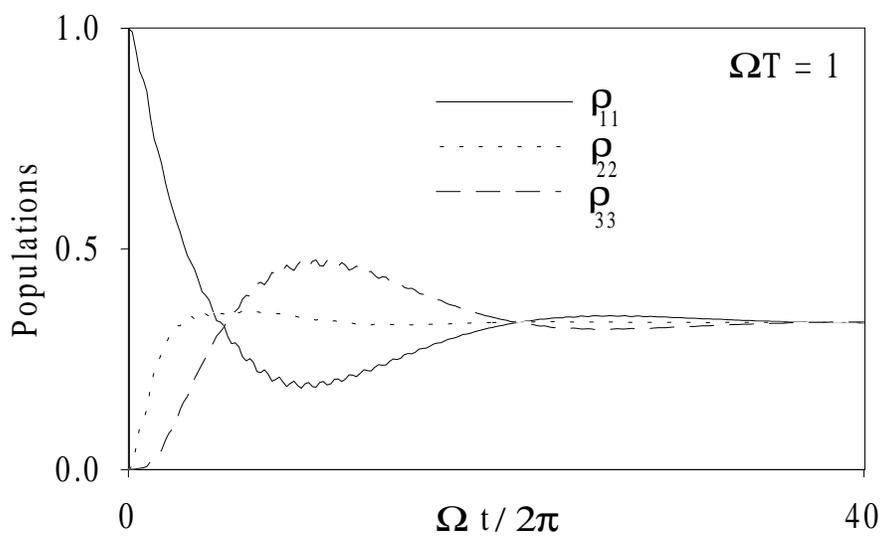

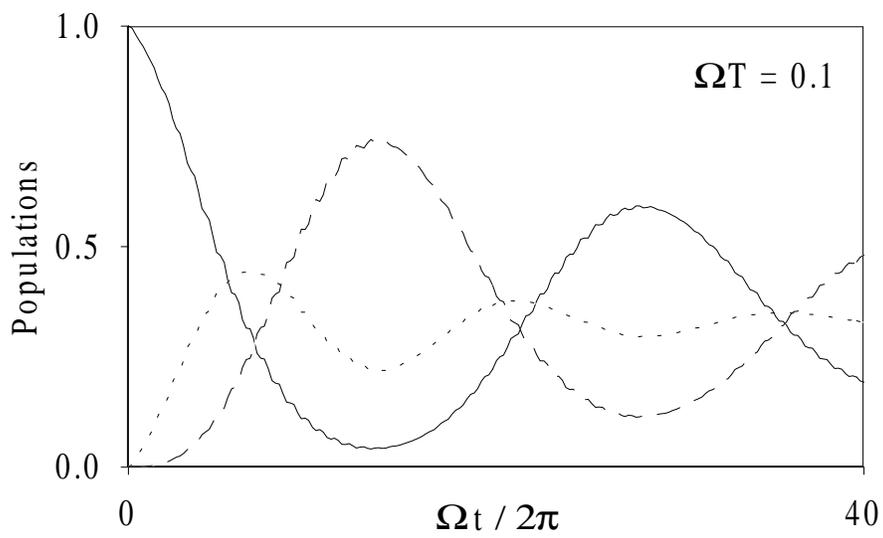

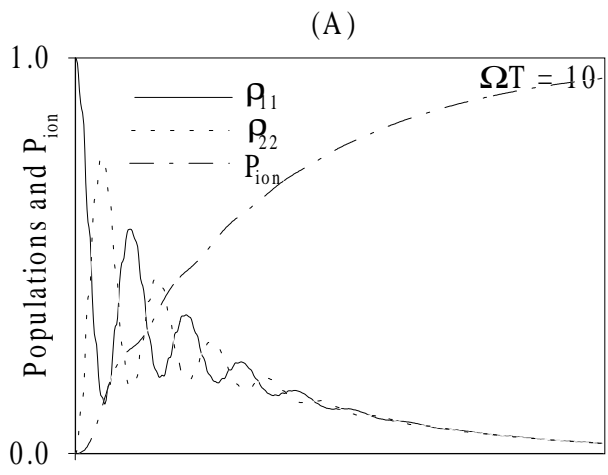
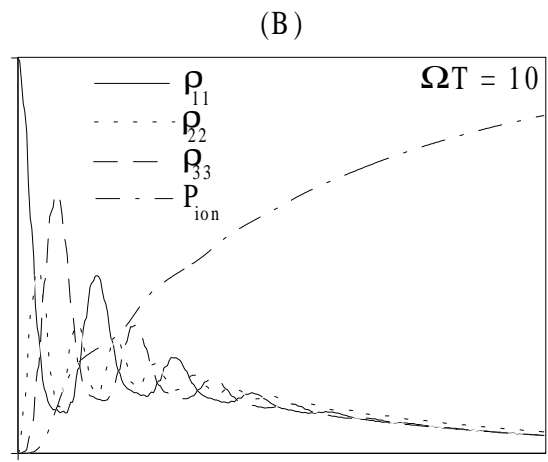
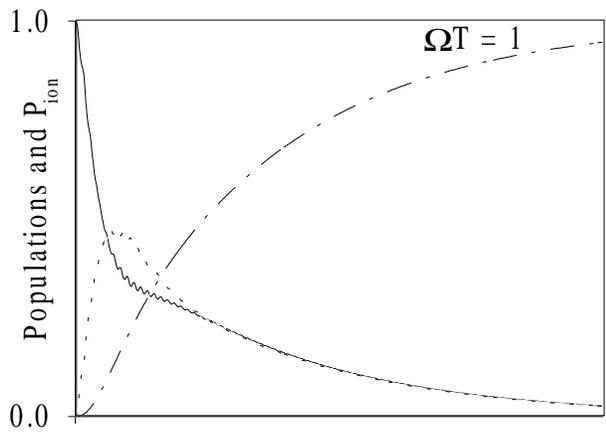
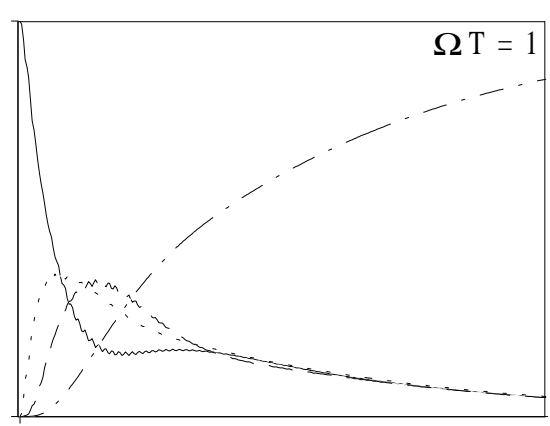
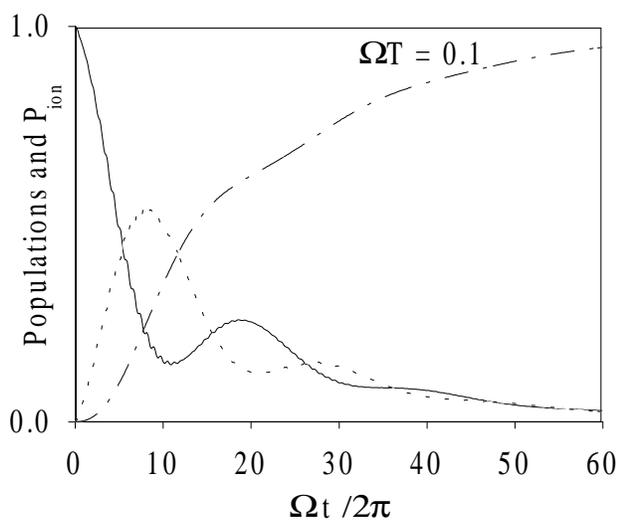
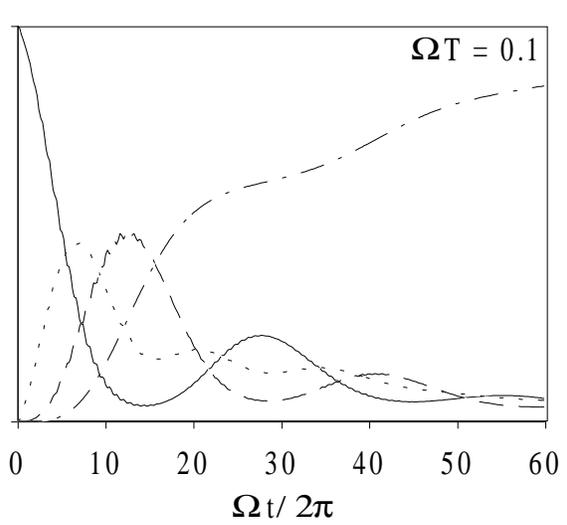

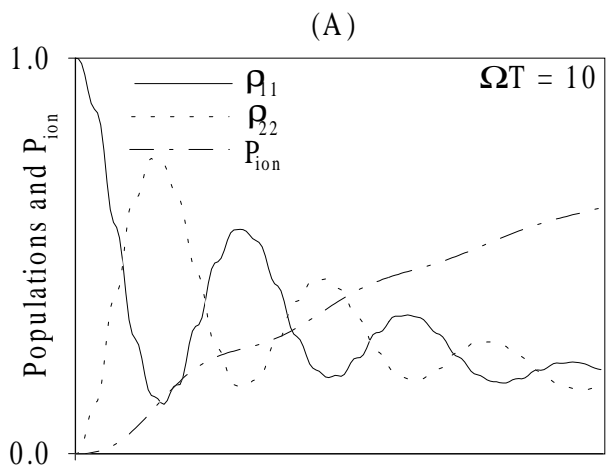
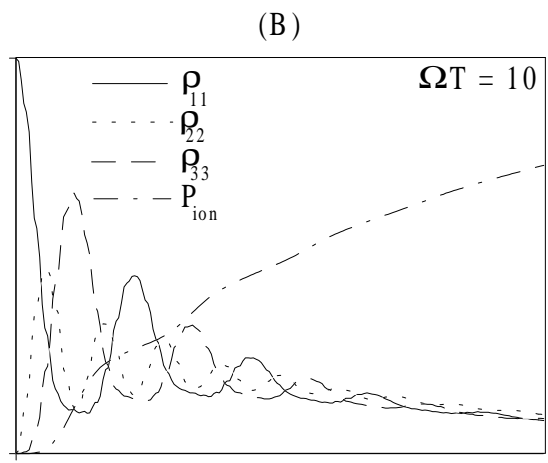
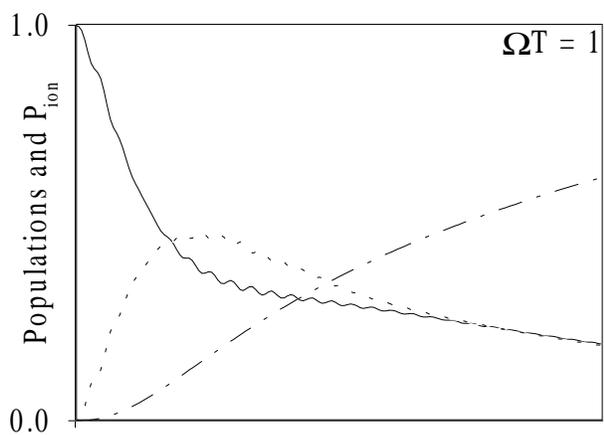
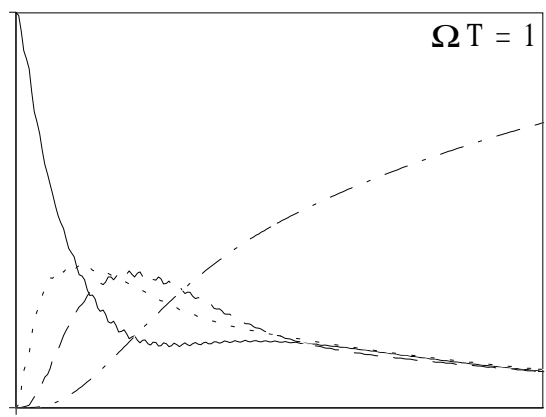
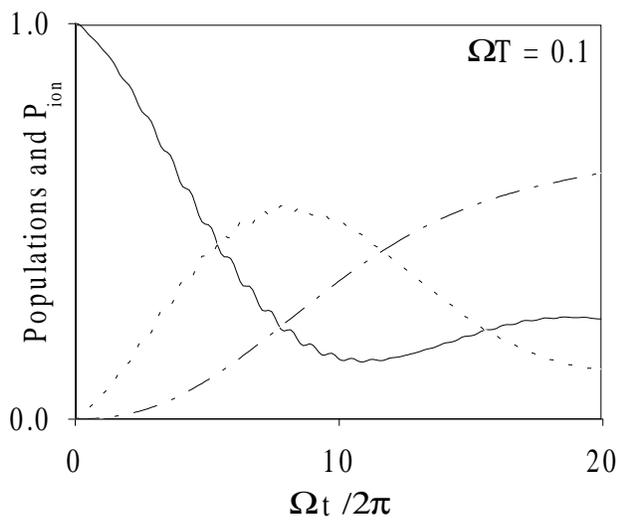
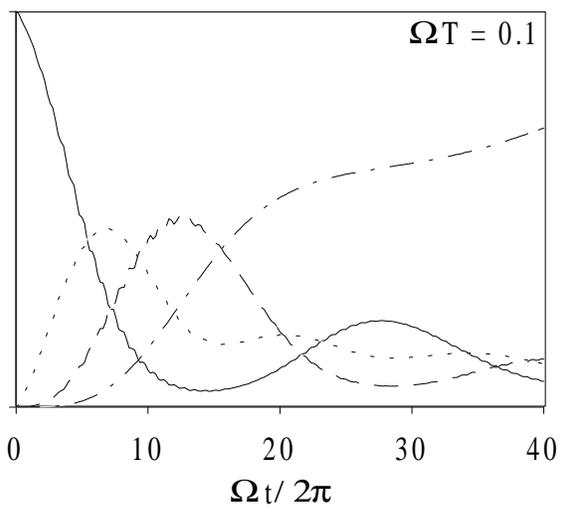

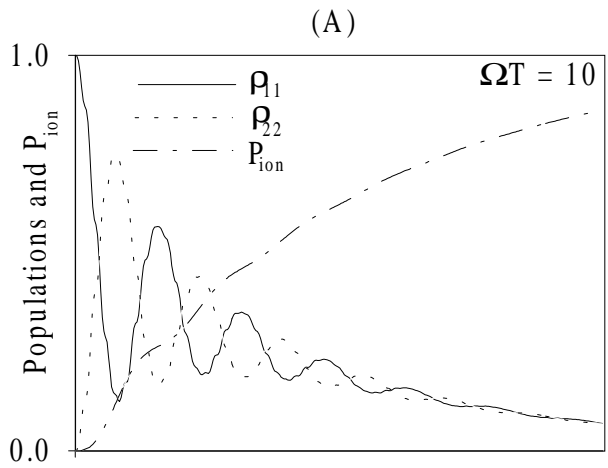
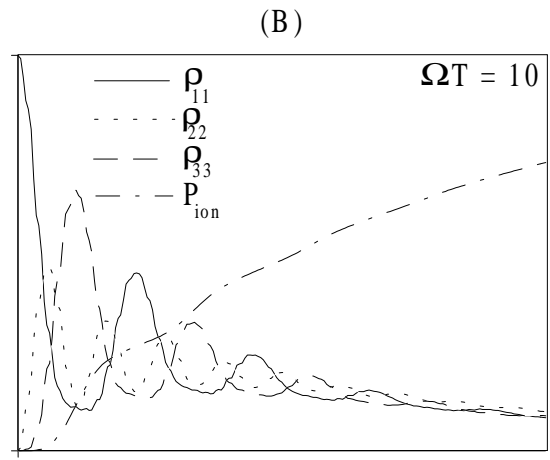
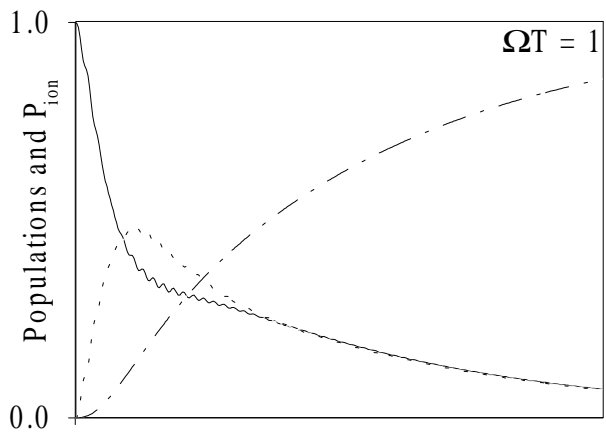
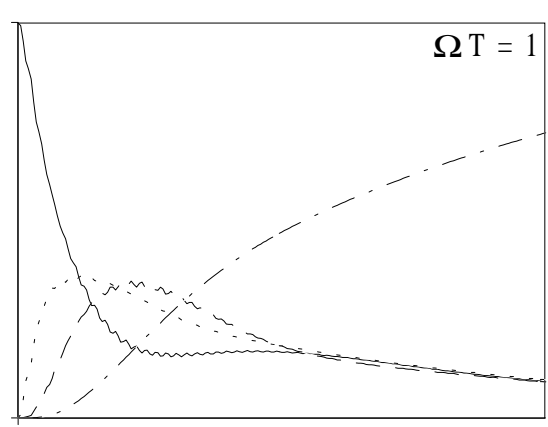
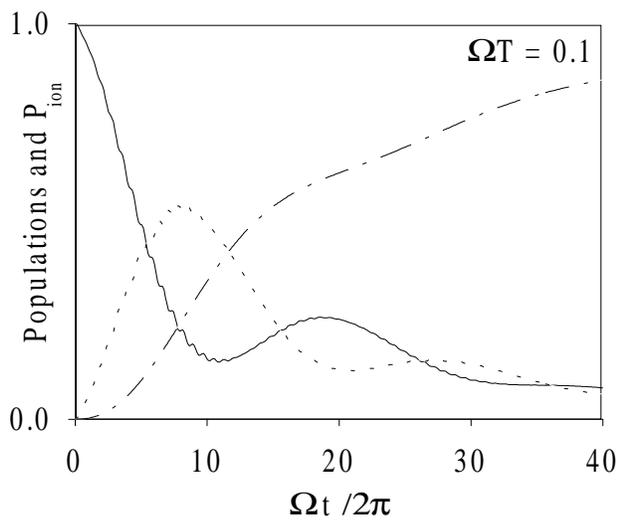
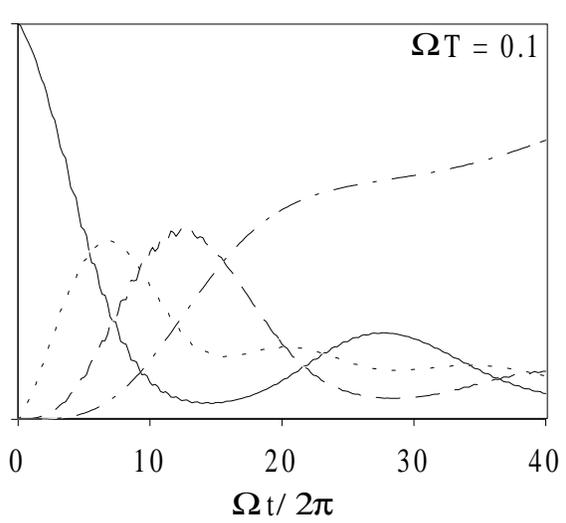